\documentclass[useAMS,usenatbib,usegraphicx]{mn2e}

\title[IFU stellar kinematics of active galaxies]{Gemini/GMOS IFU stellar kinematics of the nuclear region of six nearby active galaxies}

\author[F. K. B. Barbosa et al.]
{F. K. B. Barbosa$^{1}$, T. Storchi-Bergmann$^{1}$, R. Cid Fernandes$^{2}$,
Cl\'audia Winge$^{3}$, and
\newauthor
H. Schmitt$^{4,5}$\\
$^{1}$ Instituto de F\'{i}sica -- UFRGS, Caixa Postal 15051, CEP
91501-970, Porto Alegre, RS, Brazil \\
$^{2}$ Departamento de F\'{i}sica, CFM -- UFSC, Campus
Universit\'ario -- Trindade, CP 476, CEP 88040-900,
Florian\'opolis, SC, Brazil \\
$^{3}$ Gemini Observatory, Casilla 603, La Serena, Chile \\
$^{4}$ Remote Sensing Division, Code 7210, Naval Research Laboratory, 4555
Overlook Avenue, SW, Washington, DC 20375, USA \\
$^{5}$ Interferometric Inc., 13454 Sunrise Valley, Suite 240,
Herndon, VA20171, USA
}

\begin{document}

\date{in original form 2005 October 11}

\pagerange{\pageref{firstpage}--\pageref{lastpage}} \pubyear{2002}

\maketitle

\label{firstpage}

\begin{abstract}

We present two-dimensional (2D) mapping of the stellar velocity
field within the inner 5\arcsec\ of six nearby active galaxies,
using spectra obtained with the Integral Field Unit of the GMOS
instrument at the Gemini North telescope. 
The sampling of the observations is 0\farcs 2, corresponding at the
galaxies to spatial extents ranging from 10 to 30\,pc.
The spatial resolution range from 20 to about 180\,pc, and the
observed field of view covers a few hundred parsecs around the
nuclei.
The Calcium\,II triplet absorption features at $\approx$ 8500\AA\
were used to measure the stellar radial velocities and velocity
dispersions.
The radial velocity fields are dominated by rotation in all
galaxies.
A simple kinematical model assuming a purely rotating system with
circular orbits in a plane was fitted to the radial velocity data.
The turnover of the rotation curve is at only $\approx 50$\,pc for
NGC\,4051 and between 200 and 700\,pc for the other 5 galaxies.
The velocity dispersion ($\sigma$) maps show the largest values
($100\ge \sigma \ge 150 $ km\,s$^{-1}$) at the centre.
In the cases of NGC\,2273 and NGC\,3227, there is a decrease to
$\sigma \approx 70-80$ km\,s$^{-1}$ at $\approx 200-300$\,pc from
the nucleus, delineating partial rings of low $\sigma$ values.
A similar broken ring seems to be present at $\approx 400$\,pc from
the nucleus also in NGC\,4593. 
We interpret these low $\sigma$ rings as traces of recently formed
stars that partially keep the cold kinematics of the original gas
from which they have formed.
In NGC\,3516 there is a decrease of $\sigma$ outwards with the
steepest gradient observed along the direction of the galaxy major
axis, where $\sigma$ reaches $\approx 80-90$ km\,s$^{-1}$ at
$\approx 400$\,pc from the nucleus. 

The main novelty of the present work is the unprecedented spatial
resolution reached by a 2D study of stellar kinematics of Seyfert
galaxies using an IFU. 
The few similar IFU studies available in the literature for Seyfert
galaxies have a much poorer spatial resolution and/or are restricted
to the study of emission line kinematics.

\end{abstract}


\section{Introduction}
\mbox{}

It is now widely accepted that active galactic nuclei (AGN) are
powered by the accretion of material onto a central supermassive
black-hole (SMBH).
The present paradigm for the evolution of galaxies is that all
galaxies which form bulges also form a SMBH at their centres
(\citealt{FM00}; \citealt{geb00}).
In this scenario, the active galaxies are those in which the SMBH is
presently accreting material from its surroundings. 

A problem still under investigation is the mechanisms by which
the material is dragged to the nuclear regions
to feed the AGN, as the galactic gas must loose almost all of its
angular momentum in order to reach the central few parsecs.
Signatures that such feeding is occurring include the frequent
occurrence of recent star formation around AGN, which implies the
existence of a gas reservoir close to the AGN \citep{sch99, cid01,
sto00, boi00, cid05, sto05}.
Besides the signatures of young stars observed in spectra of AGN,
kiloparsec scale kinematic studies \citep{nw95,nw96} have
suggested that Seyfert galaxies have lower mean mass-to-light ratios
than normal galaxies, which also could be due to a younger near-nuclear
stellar population \citep{oli95,oli99}.

The presence of young stars near the nucleus seems also to be the
explanation for the results of recent studies investigating the
stellar kinematics on scales of hundred of parsecs
\citep{ems01,gar99,mar03}.
These studies have found a central drop in the stellar velocity
dispersion (hereafter $\sigma$-drop) in a few galaxies with Seyfert
nuclei.
\citet{woz03} presented simulations showing that a new generation of
stars formed at the centre of the galaxy from cold material would
create such a  drop which would remain visible for hundreds of Myrs.
Nevertheless, a decrease in $\sigma$ towards the nucleus has been
recently observed also in late-type non-active galaxies
\citet{gan06}, supporting its link to recent star-formation, but not
necessarily to nuclear activity in galaxies.

Most studies available in the literature on kinematics of Seyfert
galaxies are based on long-slit observations, which are restricted
to only one axis of the host galaxies.
In order to properly probe the galactic gravitational potential, as
well as to investigate the nature and extent of the $\sigma$-drops,
it is necessary to cover a 2D region, what we have done in the
present work using the Integral Field Unit (IFU) of the Gemini
Multi-object Spectrograph (GMOS).

The power of IFU observations to map the large scale kinematics of
galaxies has been evidenced in recent studies (e.g.
\citealt{gan06}).
For AGN, only few such studies are available (e.g. \citealt{ems06} for
NGC\,1068).
Most AGN kinematic studies have focused on gas emission-lines, as it
is much easier to obtain kinematical data from emission lines than
from stellar absorption features.
Nevertheless, the gas in the nuclear region of AGNs is subject to
non-gravitational effects, such as winds and jets, and in order to
probe the gravitational potential, it is necessary to measure the
stellar kinematics.

In this work, we use the IFU-GMOS to map the stellar kinematics in
the inner few hundred parsecs of 6 of the closest Seyfert
galaxies.
The combination of the spatial resolution at the galaxies reached by
the observations (down to a few tens of parsecs), 2D coverage,
stellar kinematics, and a Seyfert sample is unique in the
literature.

This paper is organized as follows.
In Section 2 we present the criteria used to select the sample and
discuss the relevant information for each galaxy.
In Section 3 we describe the observations and reductions, in section
4 we present the methods used to analyze the data and in Section 5
we report and discuss our results.
In section 6 we present a summary of the results and our
conclusions.

\section{Sample}
\mbox{}

We chose to map the stellar kinematics using the Ca II triplet
absorption feature  around 8500\AA\ (hereafter Ca-T) due to the fact
that this spectral region is not much affected by emission lines and
the continuum of the active nucleus, and thus clearer measurements
of the stellar kinematics can be made.

The sample galaxies were selected as the closest Seyfert galaxies
for which previous spectra in the Ca-T region were available in the
literature \citep{nw95}, so that we could check, in advance, if the
absorption features to be used in the stellar kinematic measurements
were clearly present in the spectra.
In this work, we present the results for six Seyfert galaxies, whose
properties are listed in Table \ref{sample}.

Besides selecting the galaxies on the basis of their proximity and 
detectability of the stellar Ca-T, we also looked for the presence of 
strong [S\,{\sc iii}] $\lambda 9069$ line emission, as this line 
could be included in the same observational set up used to observe the 
Ca-T.
The results based on the measurements of this emission-line, such as
intensity maps and gaseous kinematics, will be presented in a
forthcoming paper.

We have also observed kinematic standard stars, to be used as
references in the measurement of the radial velocities and velocity
dispersions of the sample galaxies.

\subsection{NGC\,2273}
\mbox{}

From a 6\,cm radio map, \citet{UW84} found the nucleus to be
resolved into two distinct components separated by 0\farcs9 along
E-W.
\citet{FWM00}, using HST, resolved the central ovoid of the galaxy,
which shows two arc-like structures forming a partial nuclear ring,
with semi-major axis of $\sim 2\farcs5$, observed both in a
broad-band and in a [N\,{\sc ii}]$+$H$\alpha$ emission-line image.
They suggest that emission in the ring, also weakly observed in
[O\,{\sc iii}], comes from H\,{\sc ii} regions.

The partial ring is also detected in a color map $\log \left(
\mathrm{F547M}/\mathrm{F791W}\right)$ which presents lower values to NW
suggesting this is the near side of the ring.
The analysis of the [O\,{\sc iii}] and [N\,{\sc ii}]$+$H$\alpha$
emission line images reveals a jet-like structure 
which extends by 2\arcsec\ to E of the nucleus and
is aligned with the radio structures observed in 6 cm.

Subsequent investigation \citep{ES03} based on HST images and colour
maps led the authors to classify the nuclear ring as a star-forming
ring.
They also classified the structure internal to the ring as a luminous
two-armed blue spiral in opposition to a previous bar-like
interpretation \citep{MR97}.

The large $H-K$ colour index and high 10\,$\mu$m luminosity,
extended by more than 5\arcsec\, also supports the presence of a
circumnuclear starburst \citep{dev89,YD91}.

\subsection{NGC\,3227}
\mbox{}

This galaxy forms an interacting pair with the elliptical galaxy NGC
3226.
\citet{GD97} reported traces of young stellar population in the
nuclear optical spectrum.
\citet{mun95} studied the H\,{\sc i} emission and found that the
galaxy disk is inclined by 56\degr with major axis at PA$=158\degr$ and
contains an  H\,{\sc i} mass of $5.7 \times 10^8$ M$_\odot$.
H$_2$ maps have been obtained by \citet{qui99} from HST/NICMOS
images and show elongated emission along PA$\sim 100\degr$.
\citet{mei90} obtained $^{12}$CO (1-0) maps showing a nuclear double
peak aligned roughly E-W separated by 2\arcsec\ and extended
emission running from SE to NW at 30\degr\ from the major axis of the galaxy
(the direction pointing to the companion galaxy NGC\,3226).
To explain these maps \citet{fer99} proposed a gas ``disk'' of 100
pc diameter with a major axis along PA$\sim 100\degr$.
Better resolution $^{12}$CO (1-0) and $^{12}$CO (2-1) maps were
obtained by \citet{SET00} who find that the central region is resolved into an
uneven ring-like structure with radius $\sim 3$\arcsec\ and the E
part six times brighter than the W part.
These maps also show the SE-NW extended structures and a nuclear bar
that connects the ring to the external NW component.
The kinematical analysis led them to conclude that a warped gas disk
provides a better description of the observed gas motions than a
bar.

\subsection{NGC\,3516}
\mbox{}

This was the first Seyfert galaxy with detected line variability
\citep{AS68}.
More recently \citet{WGY01} observed variability in both the X-ray
continuum and K$\alpha$ line.
\citet{RM99}, using HST images showed that a single dust spiral
pattern dominates the nuclear morphology.
Using {\it J}-band images \citet{qui99} proposed that this galaxy has two
bars.
\citet{nag99} found a radio jet along P.A. $\sim 10^\circ$ which
corresponds to the near side of the galaxy, according to
\citet{MGT98}, so this jet would be projected against the near
side of the galaxy.
H$\alpha\ +$ [N\,{\sc ii}] and [O\,{\sc iii}] images show a
``Z''-shaped circumnuclear emission extended by $\sim 20$\arcsec\
from SW to NE \citep{pog89, MWP92}.
The gas velocities inside this structure cannot be reproduced by
rotation, but can be explained by a nuclear outflow \citep{mul92,
VTB93}.
The galactic disc does not contain H\,{\sc ii} regions, according to
\citet{gon97}.

\citet{arr97} presented 2D stellar kinematics over a similar field
to ours but with poorer spatial resolution.
Details about their data and a comparison to ours are discussed in
Sec.\ \ref{resultados}.

\subsection{NGC\,4051}
\mbox{}

Using [O\,{\sc iii}] emission-line archival HST images, \citet{SK96}
found an unresolved nuclear source and a low surface brightness
component extending 1\farcs 2 from the nucleus along PA $=100^\circ$,
approximately the same orientation as that connecting the two radio
components at 6\,cm detected by \citet{UW84}, which are separated by
0\farcs 4.
In radio 6 and 20\,cm but with a resolution of 1\arcsec\
\citet{HU01} detected a larger scale component extending to SW and NE
from the nucleus.
\citet{vei91} reported blue wings in the forbidden optical emission
lines and proposed a model with outflow and obscuring dust.
The mid and far infrared fluxes have been explained by \citet{rod96}
and \citet{CV99} as due to hot dust emission.
In X-rays \citet{law85} found variability on time scales of hundreds
of seconds and \citet{sal93} reported a flux change by a factor 2 at
2.2~$\mu$m in 6 months.
Using X-ray maps of the nuclear region \citet{sin99} proposed the
existence of a component associated with the nuclear activity and
other associated with a starburst extended by $\sim 40\arcsec$.

\subsection{NGC\,4593}
\mbox{}

This is a Seyfert\,1 galaxy with detected nuclear variability in
X-rays by \citet{WGY01} and in the optical and infrared by
\citet{win92}, \citet{KWW93} and \citet{KW94}.
The H$\alpha$ map obtained by \citet{gon97} shows ionized gas in a
nuclear halo with radius 2-3\arcsec\ elongated N to S, and
at the border of the halo some patches of emission 
which they suggest to be a broken starburst ring with minor
axis length $\sim 5\farcs 7$ running S to N.
There are a number of H\,{\sc ii} regions in the spiral arms but
none in the large scale bar \citep{eva96}.

\subsection{NGC\,4941}
\mbox{}

According to the hard and soft X-rays spectra, \citet{MRS99}
calculated that the nucleus of this galaxy has a column density of $N_H = 4.5
\times 10^{23}$\,cm$^{-2}$.
Extended [O\,{\sc iii}] emission \citep{pog89} shows an halo
shape extending up to $\sim 10$\arcsec\ from the nucleus 
while H$\alpha$ emission is
concentrated in H\,{\sc ii} regions along the spiral arms.
\citet{sch01} found two resolved radio components
separated by 15\,pc along PA~$=335^\circ$.

\section{Observations and reductions}
\mbox{}

\label{observacoes}

\begin{table}
{\centering
\caption{Sample galaxies (column 1), morphological/Seyfert type
(columns 2/3), adopted spatial scale (column 4) and seeing measured
from the acquisition images (columns 5/6).}
\label{sample}
\begin{tabular}{@{}cccccc@{}}
\hline
Galaxy & Morph. & Seyfert & Scale & \multicolumn{2}{c@{}}{Image quality$^{\mathrm{a}}$} \\
       & type      & type & pc/(\arcsec) & (\arcsec) & pc \\ \hline
NGC 2273 & SBa\hspace{\stretch{1}}     & 2   & 120 & 1.00 & 120 \\
NGC 3227 & SAB pec\hspace{\stretch{1}} & 1.5 & \ \,84  & 0.80 & \ \,67 \\
NGC 3516 & S0\hspace{\stretch{1}}      & 1.5 & 183 & 0.98 & 180 \\
NGC 4051 & SABbc\hspace{\stretch{1}}   & 1   & \ \,45  & 0.50 & \ \,23 \\
NGC 4593 & SBb\hspace{\stretch{1}}     & 1   & 174 & 0.49 & \ \,85 \\
NGC 4941 & Sab\hspace{\stretch{1}}     & 2   & \ \,72  & 0.40 & \ \,29 \\ \hline
\end{tabular}
}
\noindent

{$^{\mathrm{a}}$
Although this has been obtained from the PSF measured in the
acquisition image, we verified that the nuclear PSF measured in the
reconstructed image of the Seyfert 1 galaxies is the same for
NGC\,4051 and only $\sim 10$\% larger for NGC\,4593.
}
\end{table}

The data were obtained in queue mode over three semesters at the Gemini
North telescope, using the GMOS IFU.
The instrumental setup for each program is shown in Table
\ref{setup}.
The spectral resolution was $R \sim 3000$ (FWHM $\sim 100$ km s$^{-1}$)
with a wavelength sampling of 0.692 \AA/pix.
The GMOS IFU consists of an hexagonal array of 1000 lenslets which
send the light through optical fibers to the spectrograph.
The lenslet array samples a field-of-view (FOV) of 7$\arcsec \times
5\arcsec$.
The sky is sampled with a field which is displaced by 1 arcmin from
the object and has one half of the size of the object field.
The centres of contiguous lenses are separated by 0\farcs2, which is also
the distance between opposite faces of the hexagonal lenses.
The effective slit aperture for every lens is, however, 0\farcs31.

Each half of the fibres (corresponding to half of the FOV) is
aligned to a separate pseudo-slit.
The data can be obtained using both slits (full FOV) or one slit
(half the FOV, or $3\farcs5 \times 5\arcsec$) with twice
the wavelength coverage.
The spectra are projected into an array of three $2048 \times 4608$
EEV chips disposed side by side with small gaps in between.
The peaks of contiguous spectra are separated by $\sim 5$ pixels.
In two-slit mode each pseudo-slit illuminates one side of the array,
and the use of passband filters avoids spectral overlap.

\begin{table*}
\centering
\caption{Details of the observations.}
\label{setup}
\begin{tabular}{@{}c@{}c@{}c@{}cc@{}ccc@{}}
\hline
PROGID        & Date    & Observed & Field size  & Exp Time & Grating/ & Spec. coverage \\
& {\footnotesize (MM/YYYY)} & galaxies & (\arcsec)& (s)    & Filter & (\AA)  \\ \hline
GN-2002B-Q-15 & 12/2002 & NGC 3227 & 3.5$\times$5 & 1440 & R400/RG610\_G0307 & 6400 - 10400 \\
              &         & NGC 3516 &              & 1440 & R400/RG610\_G0307 &              \\
              &         & BD +31 22 14  \\
GN-2003A-Q-20 & 03/2003 & NGC 4051 & 7$\times$5   & 1800 & R400/CaT\_G0309 & 8250 - 9450  \\
              &         & NGC 4941 &              & 1800 & R400/CaT\_G0309 &              \\
              &         & BD +21 24 25 \\
GN-2004A-Q-1  & 02/2004 & NGC 2273 & 7$\times$5   & 3600 & R400/CaT\_G0309 & 7700 - 9500  \\
              &         & NGC 4593 &              & 3600 & R400/CaT\_G0309 &              \\
              &         & BD +31 22 14  \\
\hline
\end{tabular}
\end{table*}

Before the spectral observations, one direct image was obtained for
centering purposes, allowing an evaluation of the image quality.
This value, measured as the FWHM of the stellar PSFs in the field,
is listed in Table \ref{sample}.

Data reduction was accomplished using generic IRAF\footnote{
IRAF is distributed by the National Optical Astronomy Observatories,
which is operated by the Association of Universities for Research in
Astronomy, Inc. (AURA) under cooperative agreement with the National
Science Foundation.
} tasks as well as specific tasks developed for GMOS data in the
{\tt gemini.gmos} package.
The reduction process comprised trimming, bias subtraction,
flat-fielding, cosmic rays cleaning, alignment and interpolation of
the data across the chips to recover all the spectra in one frame,
extraction of the spectra (the tracing is done using flat spectra),
wavelength calibration, sky subtraction and co-addition of different
exposures.

\begin{figure*}
\centering
\includegraphics[angle=-90,scale=0.8]{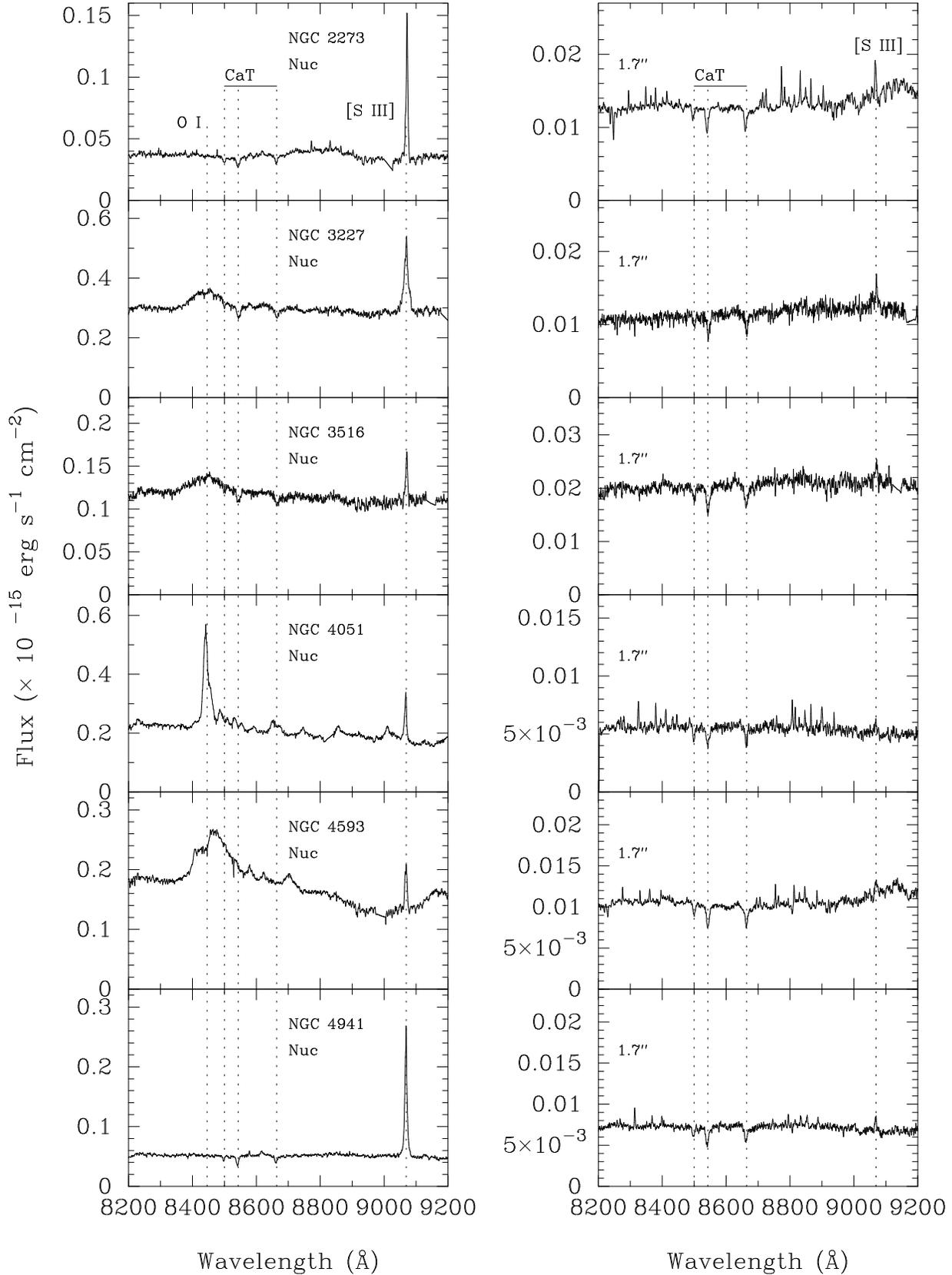}
\caption{Typical spectra of our sample.
In the left panels we show the nuclear spectra of the lens with the
strongest continuum flux (adopted as corresponding to the nucleus) 
and in the right panels the spectra of a lens
displaced 1\farcs 7 left from the centre of the array.
Each left-right spectra pair corresponds to the same galaxy and is
identified by a label in the left panel.
The spectra have been brought to rest frame and the main spectral
features are identified by vertical dashed lines.
}
\label{espectros}
\end{figure*}

In Fig. \ref{espectros} we show, for each galaxy, the spectra from
selected lenses.
The left panels show the nuclear spectra, defined as the one with
the strongest flux in the continuum around $\lambda$8500.
The right panel shows the spectra corresponding to a lens located at
1\farcs 7 from the centre of the array, towards the left side of the
IFU field.

The nuclear spectra of the Seyfert\,1 galaxies NGC\,4051
and NGC\,4593 present too much contamination from the AGN
continuum and emission lines.
In the case of NGC\,4051 these lines include the O\,{\sc i} blend at
$\lambda 8446$, Ca-T in emission and a number of Paschen emission lines.
In the case of NGC\,4593 the O\,{\sc i} emission is very broad and
there are a number of other emission lines (e.g. N\,{\sc i}~$\lambda
8703.2$).
These strong emission lines precluded the measurements of the
stellar kinematics within $\sim 0\farcs 8$ from the nucleus in these
galaxies.
This effect was unexpected on the basis of the previously inspected
integrated spectra \citep{nw95}, probably because it was diluted in
their large aperture.
In any case, the information we could obtain for these galaxies
outside the contaminated region is still valuable and have been
included in the present work.
The spectra of the Seyfert\,1.5 galaxies NGC\,3227 and NGC\,3516
(some authors classify the latter as Sy\,1) present weak
contamination from the broad O\,{\sc I} line which, however, does not
affect the measurements significantly, as the two strongest
absorption lines of the Ca-T are unaffected.

Even after subtraction of the sky contribution, there remained some
sky lines residuals in the extranuclear spectra, observed in Fig.
\ref{espectros}.
For these lines as well as for the faint emission lines in the
nuclear spectra of NGC\,4941, we have set the parameters of the {\it
xcsao} task (used in the velocity dispersion measurements) to
interpolate the continuum eliminating all data points with fluxes
higher than 1 standard deviation of the average continuum flux
within the cross-correlation window.
We have checked that this procedure efficiently eliminated the
residual emission lines.

\section{Data analysis}

\subsection{Velocity measurements; cross-correlation technique}
\mbox{}
\label{velocitymeasurements}

The kinematic measurements were performed by cross-correlating the
spectra of individual lenslets with the spectrum of a kinematic
standard star observed with the same setup as the galaxies.
The cross-correlation was performed using the task {\tt xcsao} of
the package {\tt rvsao} \citep{KM98} in IRAF, over the spectral range
8430-8900 \AA\ which includes the Ca-T features but avoids the noisy
regions at the extremes of the spectra.
The task uses the method of the quotient of the Fourier
transforms applied to the spectra in velocity space and
searches for the peak of the cross-correlation function, fitting
a quartic function to the data with values above 50\% of the
peak and giving as outputs the peak velocity and the full width at
half maximum (FWHM) of the cross-correlation function.

As the surface brightness of the galaxies decreases towards the
borders of the field, in order to improve the signal-to-noise ratio 
at these locations, we replaced each
individual spectrum by the average of itself and the 6 nearest
spectra (which are closer than 0\farcs 3).
This replacement was done for the spectra beyond a radius 1\arcsec\ 
from the centroid of the 8500\AA\ continuum brightness
distribution.

In order to obtain the velocity dispersion values from the measured
line widths we have convolved the spectrum of the kinematic standard
star with Gaussian curves of known FWHM -- where FWHM $=
2.35\,\sigma$, and $\sigma$ is the velocity dispersion -- to create
a set of synthetic spectra.
We then performed the cross correlation between these synthetic spectra
and that of the kinematic standard, measuring the FWHM of the cross
correlation function as we did for the galaxies.
By plotting the measured widths against the known Gaussian widths we
obtain a very tight linear relation which is then used
to obtain the real FWHM from the measured FWHM for each galaxy.
Finally, as the radial velocities are measured relative to the
kinematic standard star, the radial velocity obtained for each
galaxy was corrected by the observatory motion relative to the local
standard of rest and by the standard star radial
velocity as determined from the shifts between the measured
wavelengths of the Ca-T lines in the standard spectrum and their
rest wavelengths.

\subsection{Error calculation}
\mbox{}

The IRAF cross-correlation task {\it xcsao} calculates the error
following the method of \citet{TD79}, which is similar to that used
by \citet{nw95}.
This method uses the signal-to-noise ratio $R$ and the
FWHM $w$ of the cross-correlation peak, assuming that the error in
$\sigma$ is the same as  in $V_r$:

\begin{equation}
error = \Delta \sigma = \Delta V_r \simeq \frac {3 w}{8 (1 + R)}.
\end{equation}

For the spectra where the Ca-T lines are clearly detected and are not
contaminated by emission lines, we obtain errors in the range
4-17\,km\,s$^{-1}$ for all galaxies.
However, there is a loose correlation between the velocity
dispersion measurements and the error calculated by this formula,
due to a dependence in $w$ that is not fully canceled by $(1 + R)$,
an effect also pointed out by \citet{nw95}.
This led us to adopt a conservative approach regarding the error
reported by the {\it xcsao} task.
After inspecting some representative spectra and the calculated
error values, we adopted $15$\,km\,s$^{-1}$ as the errors in our
individual measurements for $V_r$ and $\sigma$.
We note that this is actually an upper limit, which may
overestimates the actual errors for some of the measurements, in
particular the ones from the more central spectra.

\subsection{Alternative velocity measurements; pixel fitting
technique}
\mbox{}

In order to investigate the dependence of the velocity fields and in
particular of the velocity dispersions $\sigma$ on the measurement technique
we have also carried out a kinematic analysis in pixel space
\citep{bar02,cap04}.

Our implementation of the direct-fitting-method is essentially the
one used in \citet{gar05}, the only difference being that
we rectified all spectra to account for the different slopes of
stellar and galaxy spectra.

\begin{figure*}
\centering
\includegraphics[angle=-90,scale=0.9]{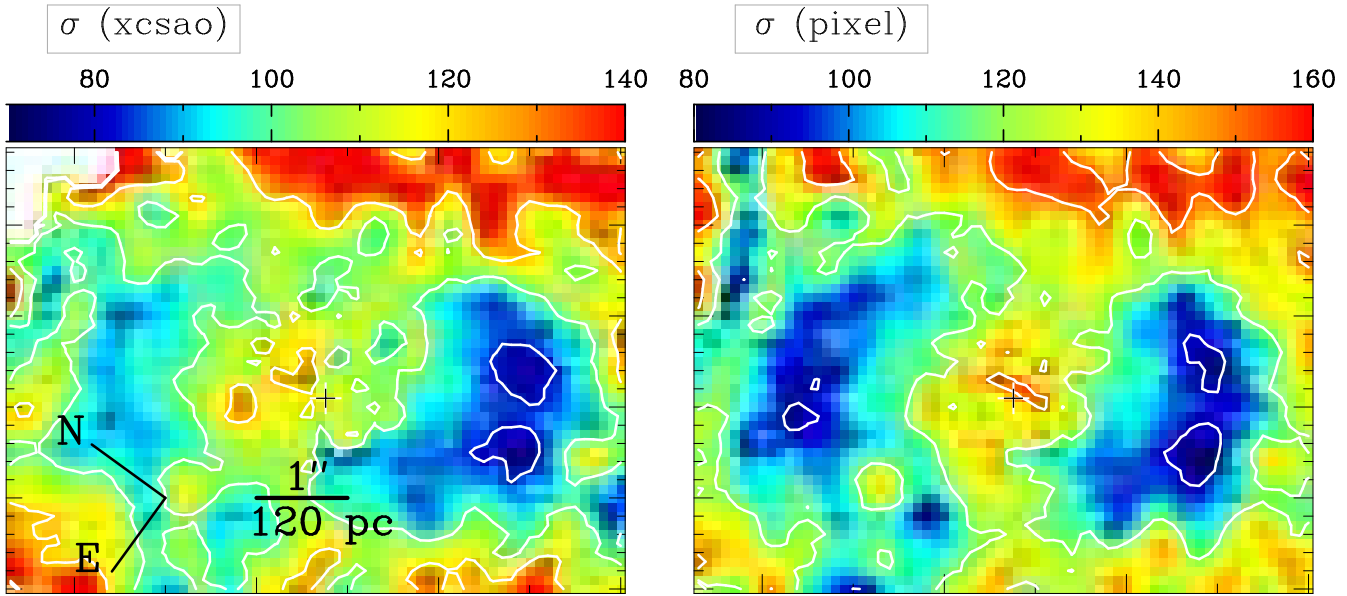}
\caption{Comparison of $\sigma$ maps obtained with different
methods for NGC\,2273.
The map on the left was obtained using the cross-correlation method, the
one to the right was obtained using a pixel space fitting technique.
In both cases we obtain the same main structure: a low $\sigma$
partial ring at $\approx 2$\arcsec\ from the nucleus.
}
\label{comp-2273}
\end{figure*}

We have compared our cross-correlation results with results obtained
with this method.
We find that the radial velocity measurements are essentially
unaltered.
We also find that the patterns in the velocity dispersion maps
remain the same.
We do find a systematic difference in the velocity dispersion values
obtained from the two methods but this difference varies from galaxy
to galaxy.
For example, for NGC\,4051 there is no difference at all while for
NGC\,2273 we find the largest difference.
This is illustrated in Fig. \ref{comp-2273}, where we compare the $\sigma$
maps obtained with both techniques for NGC\,2273.
Note that the low dispersion ring (see discussion in Section
\ref{sec-res-2273}) is clearly present in both maps,
the only difference being that the values obtained with the direct
pixel fitting method are larger by $\sim 20-30$ km s$^{-1}$.

In order to investigate if this difference could be due to a
template mismatch, we have performed velocity dispersion
measurements using both techniques and a set of 28 stellar templates
from the atlas of \citet{cen01}.
The stars were chosen to span a wide range of spectral types.
For the cross-correlation technique the results did not vary much
for template spectral types G, K and M.
Nevertheless, as we moved to earlier spectral types the results
began to vary but by simple visual inspection of these template
spectra we verified that they are bad matches to the galaxy
spectra.
In the case of the direct pixel fitting method we made two tests.
First we modeled each
spectrum with a combination of all 28 templates, letting their proportions
vary freely, thus accounting for variations in stellar populations
\citep{gar05}.
Then we repeated the fit using only our observed template star.
The difference is negligible (at most a few km s$^{-1}$),
demonstrating that our kinematic measurements are not seriously
affected by template mismatch.

\subsection{Modeling of the radial velocity field}
\mbox{}

The radial velocity maps obtained from the measurements described
above (see Fig. \ref{painel-2273}-\ref{painel-4941}) show that a
clear rotation pattern dominates the stellar kinematics.
In order to obtain an analytical description of the radial velocity
field, we have adopted a very simple approximation, assuming plane
keplerian orbits for the stars, and representing the gravitational
potential of the bulge by a Plummer potential:

\begin{equation}
\Phi = - {GM \over \sqrt{r^2 + a^2}}
\end{equation}

\noindent which depends only on the scale length $a$ and bulge mass 
$M$, and where $r$ is the radial distance in the plane of the
galaxy.

We assume that the stars are in orbits close to a plane
$P(i,\psi_0)$ characterized by its inclination relative to the plane of the
sky ($i$) and the position angle (PA) of the line of nodes $\psi_0$.
Although we do not expect that this simple approximation provides a
robust determination of the bulge mass, inclination and bulge scale length, we
can obtain reliable determination of the systemic velocity $V_s$,
PA of the line of nodes $\psi_0$ and the position of the kinematical centre.

Using the above potential, we calculate the rotation curve in the
plane $P(i,\psi_0)$ with coordinates of the origin $(X_0, Y_0)$.
The 2D rotation curve is then projected onto the plane of the sky
where the radial variable is now $R$, the angular variable is $\psi$
and the correspondent scale length is $A$.
The relations between $r$ and $R$, and between $a$ and $A$ are:
$r = \alpha R$ and $a = \alpha A$, where $\alpha = \sqrt{cos ^2(\psi -
\psi_0) + \frac{\sin ^2 (\psi- \psi_0)}{cos ^2 ( i )} }$.
The systemic velocity $V_s$ is added to the model as a zero point
offset.

In summary, there are 7 free parameters to be determined by
fitting the non-linear model

\begin{eqnarray}
\nonumber \lefteqn{V_r = V_s +}
\\
\lefteqn{
+ \sqrt
{
	    \frac
	    {R^2\,GM}
	    { \left( {R}^{2}+A^2 \right) ^{3/2}}
}
\frac
{
   \sin \left( i \right)
   \cos \left( \psi- \psi_0 \right)
}
{
   \left(
	\cos ^2 \left( \psi- \psi_0 \right)
	+
	\frac
	{
		\sin ^2 \left( \psi- \psi_0 \right)
	}
	{
		\cos ^2 \left( i \right)
	}
   \right)^{3/4}
}
}
\end{eqnarray}

\noindent to the data $V_r$, what has been done using a Levenberg-Marquardt algorithm.  

Before performing the fits, we have  inspected  the map of uncertainties in
radial velocity generated by the cross-correlation routine, the
individual spectra and the radial velocity maps in order to obtain
an uncertainty threshold to discriminate between reliable data (or measurement)
and noise.
This threshold varied somewhat from galaxy to galaxy,
ranging from 9 to 17\,km\,s$^{-1}$.
Only reliable measurements were used in the data analysis and model fitting.
The measurements considered unreliable have been masked in the
velocity maps shown in Figs. 
\ref{painel-2273}-\ref{painel-4941}.

The parameters obtained from the fits are shown in Table
\ref{parameters}.
The kinematical centre $(X_0, Y_0)$ was used to
calculate $\Delta X_0 = X_0 -X_0^{b}$ and $\Delta Y_0 = Y_0 - Y_0^{b}$
where $X_0^{b}$ and $Y_0^{b}$ are the coordinates of the centroid of
the continuum brightness distribution.

\begin{table*}
\caption{Parameters derived from our modeling.
}
\label{parameters}
\begin{tabular}{ccccccccc}
\hline
Galaxy & $V_s$     & $\Psi_0$ & $\Delta X_0$ & $\Delta Y_0$ & $M$ & $A$ & $i$ \\
       & (km/s)$^\mathrm{a}$  &  (\degr) & (pc) & (pc)    &
       ($\times10^{9} M_\odot$) & (kpc) & (\degr) \\\hline
NGC 2273 & 1836 & 233.4 & -7 $\pm$ 12 & -20 $\pm$ 12 & 2.4 & 0.17 & 51 \\
NGC 3227 & 1201 & 149.7 & -2 $\pm$ 9  & -10 $\pm$ 9 & 1.4 & 0.11 & 43 \\
NGC 3516 & 2698 & 48.4 & -18 $\pm$ 18  & 9 $\pm$ 18 & 15 & 0.28 & 22 \\
NGC 4051 & 718 & 106.5 & 4 $\pm$ 5 & 8 $\pm$ 5 & 0.084 & 0.031 & 31 \\
NGC 4593 & 2531 & 88.5 & 1 $\pm$ 17  & 4 $\pm$ 17 & 18 & 0.41 & 25 \\
NGC 4941 & 1161 & -0.2 & 1 $\pm$ 7  & -14 $\pm$ 7 & 3.6 & 0.24 & 24 \\
\hline
\end{tabular}

\medskip
\flushleft{$^{\mathrm a}$ The upper limit for the error in $V_s$ 
is about 15\,km\,s$^{-1}$.\\
$^{\mathrm b}$ $\Delta X_0$ and $\Delta Y_0$ are measured
relative to the continuum centroid (see text).}

\end{table*}

\section{Results}
\mbox{}
\label{resultados}

In this section we present and discuss the kinematic measurements 
obtained from the spectra as well as the results of the modeling.
 
\begin{figure*}
\centering
\includegraphics[angle=-90,scale=1.5]{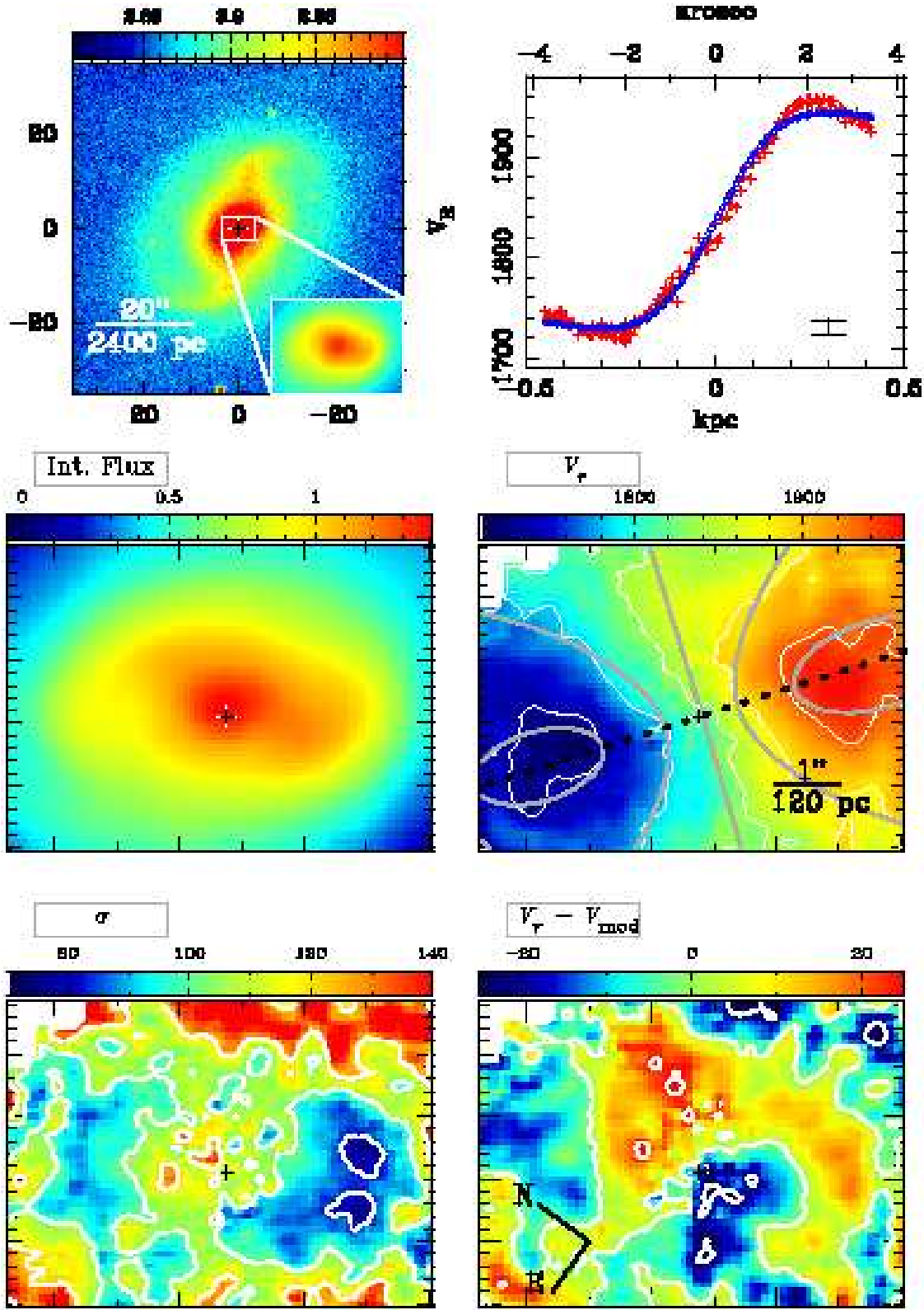}
\caption{Kinematic data for NGC\,2273.
{\it The upper left panel} shows a large scale i$^{\prime}$-band image
of the galaxy (spatial units are arcsec) with the IFU field
overlaid as a white box and an inset detailing the nuclear region
within the FOV of the IFU maps.
{\it The central left panel} shows the flux integrated over
the wavelength range of the cross-correlation measurements.
{\it The bottom left panel} shows the velocity dispersion map.
{\it The upper right panel} shows the measured radial velocities
from a virtual slit placed along the line of nodes (red), together
with the corresponding model values in (blue).
{\it The central right panel} shows the radial velocity map with
isovelocity contours from data and model superimposed (white for the measurements and
grey for the model); the cross marks the kinematic centre and the
dashed line shows the line of nodes. {\it The bottom left panel} shows the velocity dispersion map.
{\it The bottom right panel} shows the residuals between the
measured radial velocities and the model.
All panels have the same orientation, indicated at the bottom right
panel.
The four bottom panels share the scale indicated in the central right panel, and angular extents of 6\farcs 8 $\times$ 4\farcs 9.
Regions in white correspond to uncertain measurements which have not
been included in the fits.}
\label{painel-2273}
\end{figure*}

\begin{figure*}
\centering
\includegraphics[angle=-90,scale=1.5]{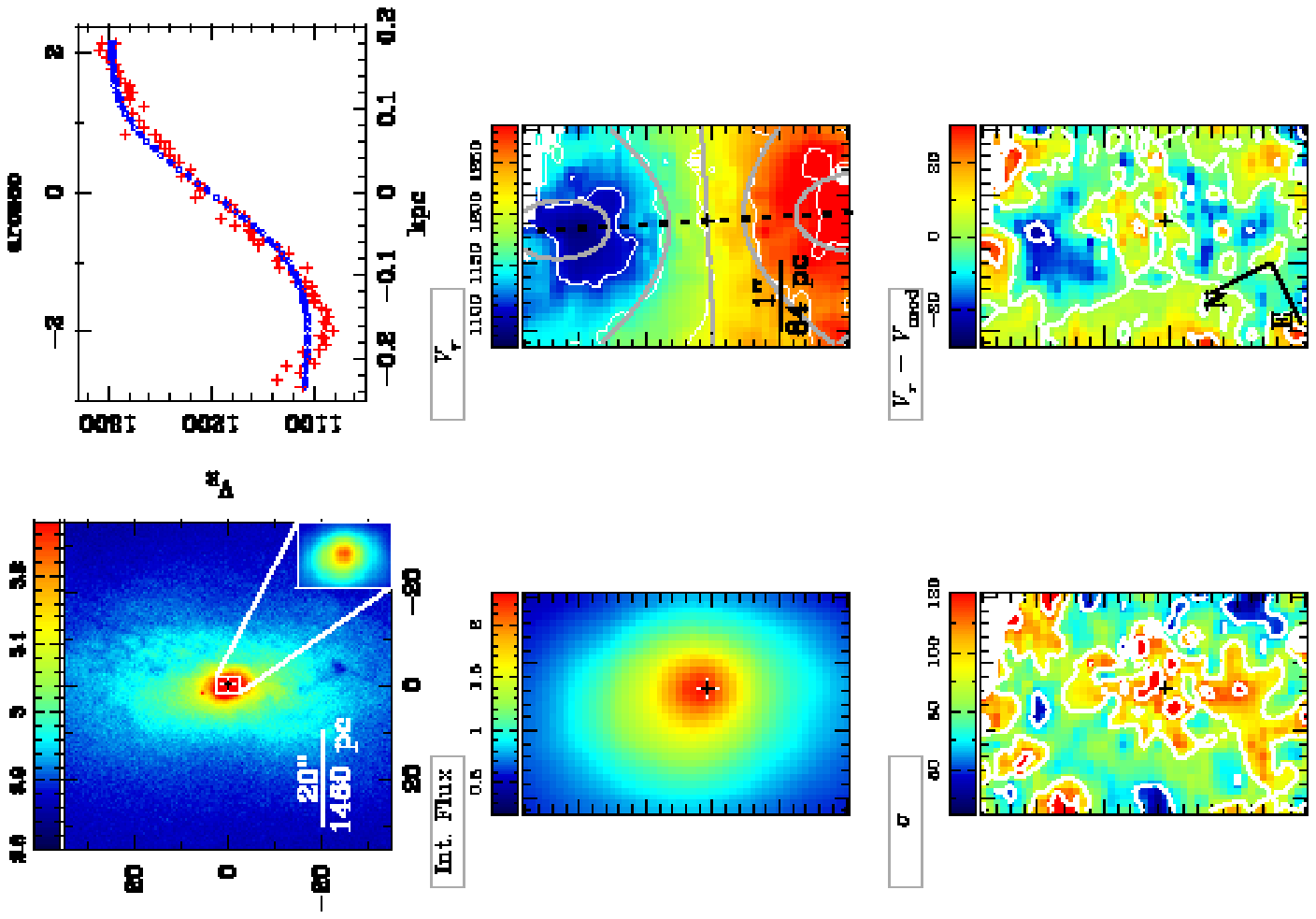}
\caption{Kinematic data for NGC\,3227.
Details are as described in the caption of Fig. \ref{painel-2273},
except for the fact that the IFU field is half the size, or
3\farcs 3 $\times$ 4\farcs 9.}
\label{painel-3227}
\end{figure*}

\begin{figure*}
\centering
\includegraphics[angle=-90,scale=1.5]{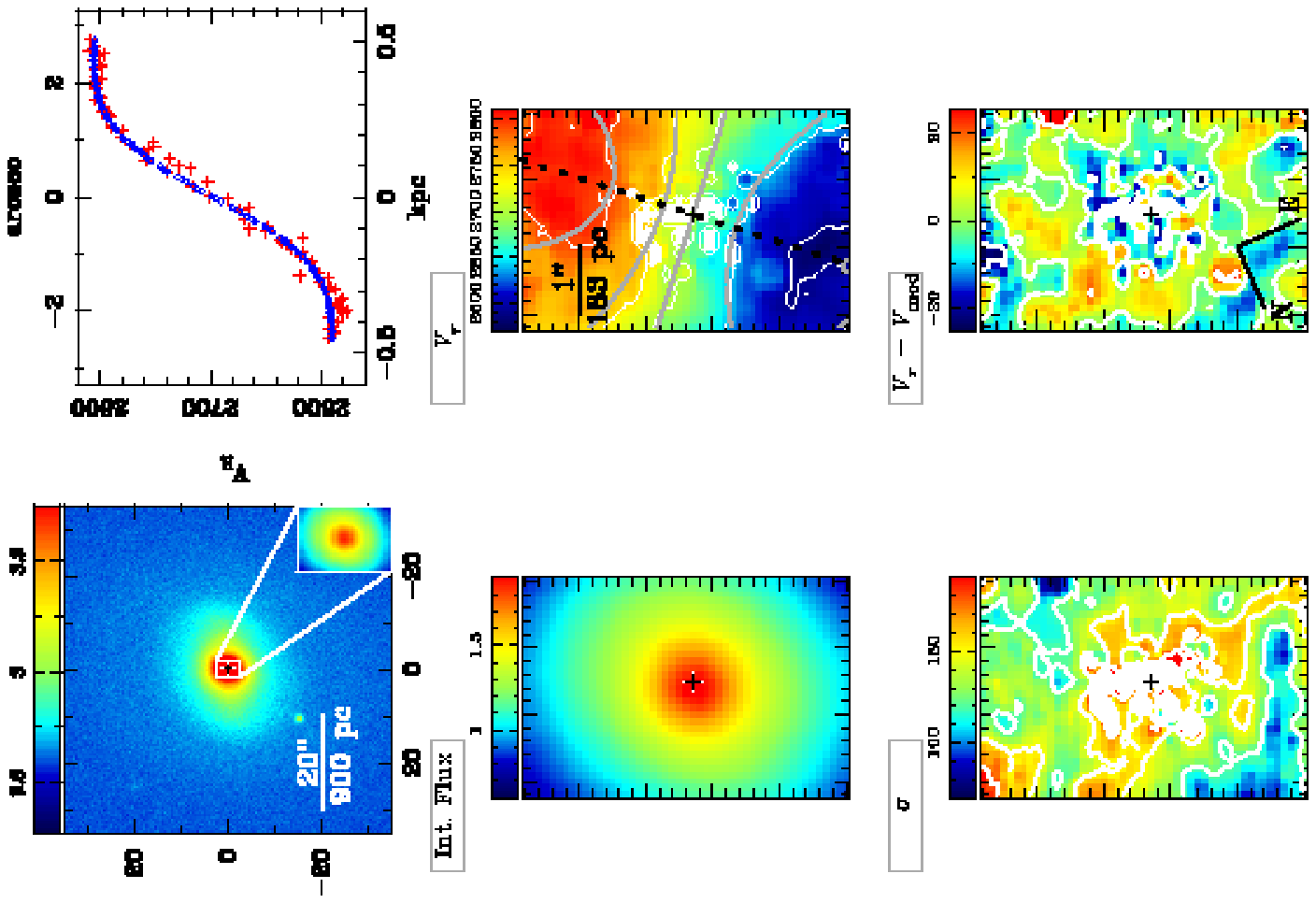}
\caption{Kinematic data for NGC\,3516.
Details are as described in the caption of Fig. \ref{painel-2273},
except for the fact that the IFU field is half the size, or
3\farcs 3 $\times$ 4\farcs 9.}
\label{painel-3516}
\end{figure*}

\begin{figure*}
\centering
\includegraphics[angle=-90,scale=1.5]{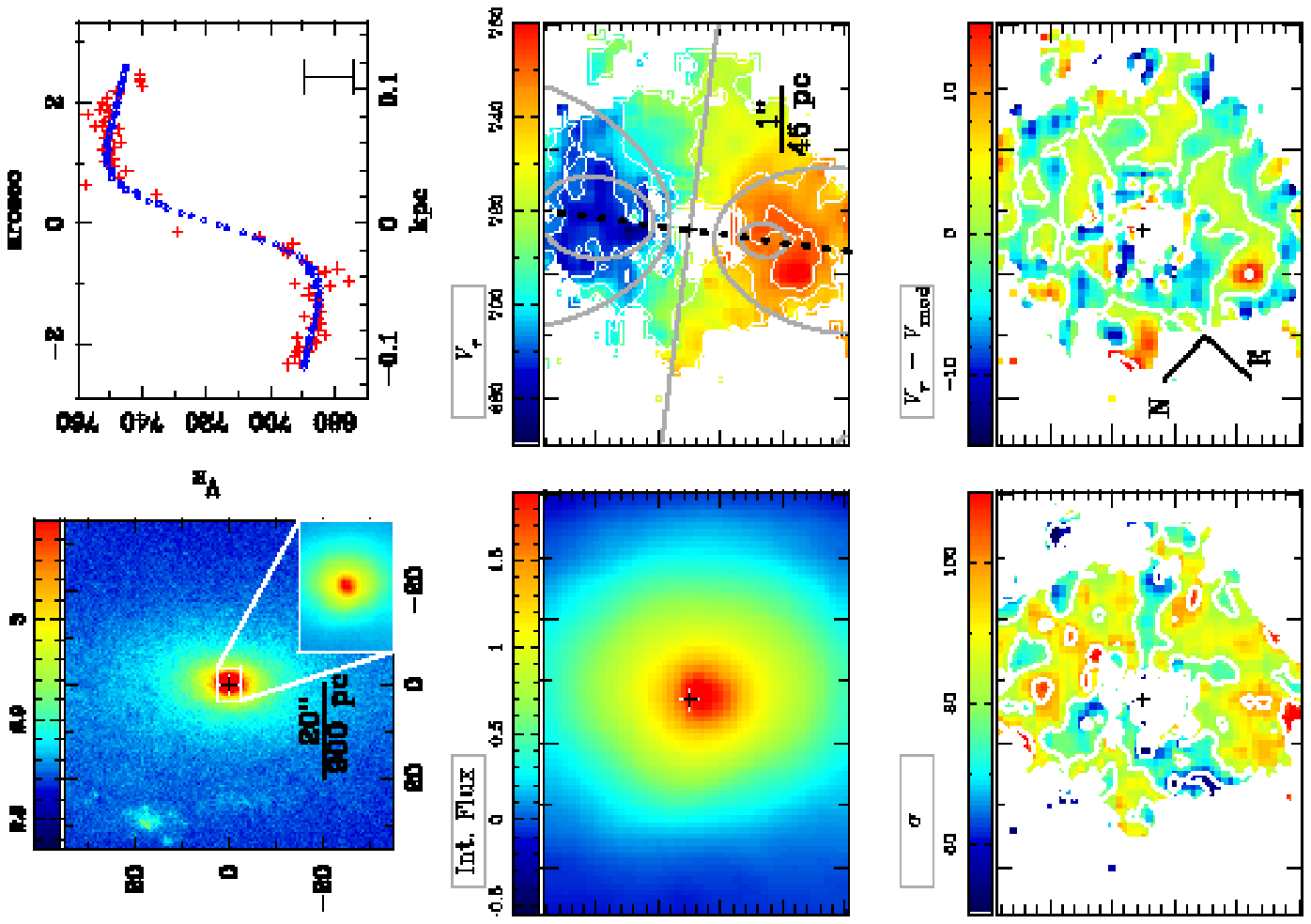}
\caption{Kinematic data for NGC\,4051.
Details are as described in the caption of Fig. \ref{painel-2273}.}
\label{painel-4051}
\end{figure*}

\begin{figure*}
\centering
\includegraphics[angle=-90,scale=1.5]{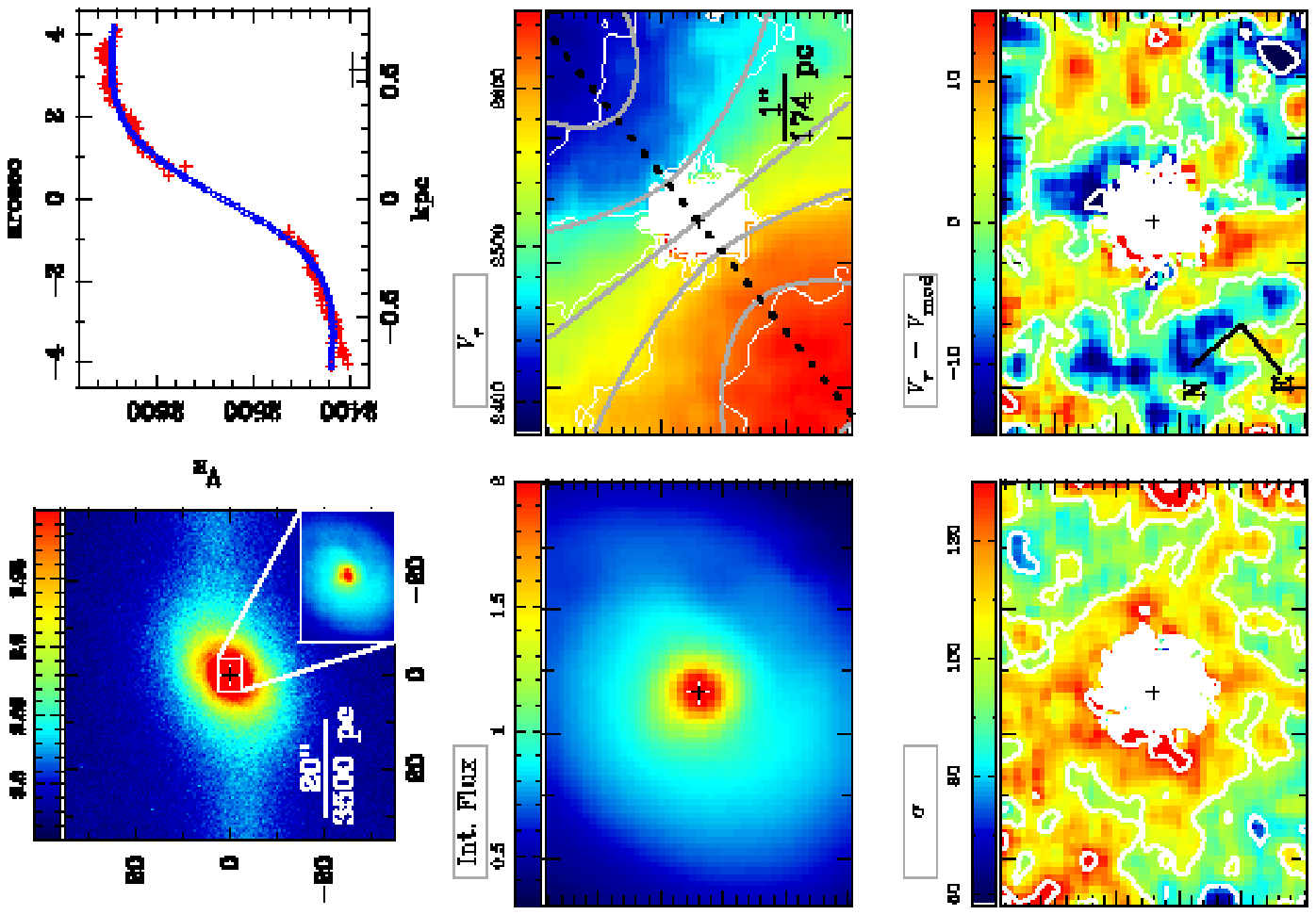}
\caption{Kinematic data for NGC\,4593.
Details are as described in the caption of Fig. \ref{painel-2273}.}
\label{painel-4593}
\end{figure*}

\begin{figure*}
\centering
\includegraphics[angle=-90,scale=1.5]{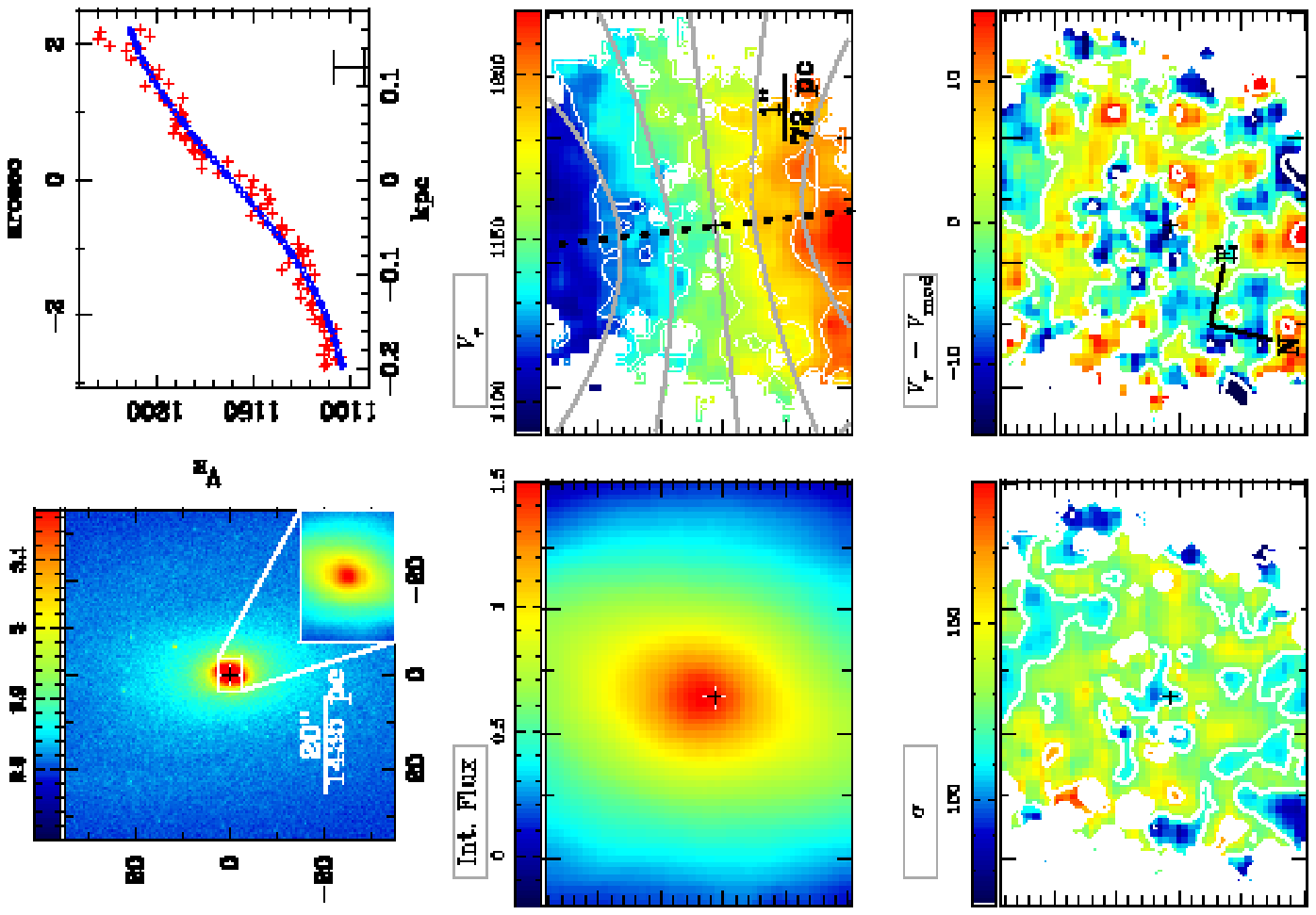}
\caption{Kinematic data for NGC\,4941.
Details are as described in the caption of Fig. \ref{painel-2273}.}
\label{painel-4941}
\end{figure*}

2D maps of the measurements, together with the acquisition
image and residuals of the model fits for each galaxy, are presented
in Figs. \ref{painel-2273}-\ref{painel-4941}.
Each figure comprises 6 panels, as described below. 

\begin{enumerate}

\item The top left panel shows a large scale {\it I}-band
image with the IFU field drawn as a white box and the peak of the
brightness distribution marked by the black cross.
The small inset at the lower right corner details the morphology of
the nuclear region covered by the IFU field.

\item The top right panel shows the measured radial
velocities $V_R$ (red crosses) along a virtual slit placed along the
line of nodes and with width 0\farcs6, together with the
corresponding model values (blue circles), as a function of the
projected distance from the nucleus\footnote{
Note that, due to the fact that the
slit width includes more than one lens there may be more than one
velocity value for each given radius at the slit, even for the
model.} $R$.
The error bar at the lower right corner represents the
maximum uncertainty in the velocity values.

\item The remaining panels show 2D maps of measured properties.
The middle left panel shows the flux integrated
over the wavelength range used in the
cross-correlation measurements.

\item The middle right panel shows the radial velocity map,
with isovelocity contours superimposed; white contours correspond to
the measurements, while grey contours correspond to the best fit model.
The kinematic centre is identified by the black cross,
and the line of nodes is indicated by the black dashed line.

\item The bottom left panel shows the velocity dispersion
$\sigma$ map. 

\item The bottom right panel shows the residuals obtained from the
subtraction of the kinematic model from the radial velocity map of
panel (iii). 

\end{enumerate}

For display purposes, the kinematic maps of Figs.
\ref{painel-2273}-\ref{painel-4941} have been interpolated over the
original hexagonal array to generate a rectangular grid uniformly
sampled, with pixel size  $0\farcs 1 \times 0\farcs 1$.
The scale (shown in panel (iii)) and orientation (shown in panel
(v))
are the same for all the maps for a given galaxy.

Table \ref{medidas} lists the velocity dispersions measured from
the integrated nuclear spectra inside circular\footnote{Actually the
combined spectra from all hexagonal elements
with centres within the circular aperture radii.}
apertures with diameters 0\farcs 6 and 1\arcsec\ for those galaxies
with reliable nuclear data, as discussed in Section~3.
Comparing our values with those from \citet{nw95}, we
conclude that they are compatible given the uncertainties 
and our smaller apertures.
The value given by \citet{tdt90} for NGC\,3516 is significantly larger
than ours, but these authors also present higher values than those of
\citet{nw95} for other galaxies in common.
For NGC\,4051 and NGC\,4593 the central velocity dispersions 
are unreliable due to the strong contribution from the AGN
continuum.

\begin{table}
\centering
\caption{Comparison between our $\sigma$  measurements and
and those from the literature for the central spectra.
Column 2 and 3 presents our measurements (see text); column 4 and 5
present the values from the literature and corresponding apertures.
The upper limit for the error in $\sigma$ is $\pm
15$\,km\,s$^{-1}$.}
\label{medidas}
\begin{tabular}{ccccc}
\hline
         & $\sigma$ (0\farcs 6) & $\sigma$ (1\arcsec) & $\sigma$
	 (NW95) & Aperture \\
         & (km/s)   & (km/s)   & (km/s)         & (\arcsec) \\\hline
NGC 2273 & 104     &  105     & 136 $\pm$ 22 & 1.5 x 1.4 \\
NGC 3227 & 114     &  109     & 128 $\pm$ 13 & 1.5 x 2.1 \\
NGC 3516 & 192     &  178     & 235$^{\mathrm a}$ \ \  \ \ \ \ & 2.1 \\
NGC 4051 & \multicolumn{2}{c}{85$^{\mathrm b}$} & \,\ 88 $\pm$ 13 & 1.5 x 2.1 \\
NGC 4593 & \multicolumn{2}{c}{105$^{\mathrm b}$} & 124 $\pm$ 28    & 1.5 x 3.5 \\
NGC 4941 & 138     &  136     & 109 $\pm$ 13    & 1.5 x 2.1 \\
\hline
\end{tabular}

\medskip
\flushleft{$^{\mathrm a}$ From \citet{tdt90}.\\
$^{\mathrm b}$ These values were obtained by inspection
of the region around the contaminated central portion of the maps.
See Sec. \ref{observacoes} for details.}
\end{table}

We now discuss the results obtained for each galaxy.

\subsection{NGC\,2273}
\mbox{}
\label{sec-res-2273}

The radial velocity map (central right panel of Fig.
\ref{painel-2273}) shows a rotation pattern which deviates from the
classical spider diagram, evidenced by the fact that the observed
kinematic minor and major axes are not perpendicular to each other, 
indicating deviations from axial symmetry.
The residuals map show indeed deviations from the model above the
noise level, which seem to be co-spatial with the star-forming ring (see 
discussion below).

The rotation curve (upper right panel of Fig. \ref{painel-2273}) 
along the kinematic major axis seems to have 
reached the turnover at only 250\,pc from the nucleus.
In addition, the data turnover seems to be more pronounced than that
of the model, suggesting a more concentrated mass distribution.

\begin{figure}
\centering
\includegraphics[angle=-90,scale=0.45]{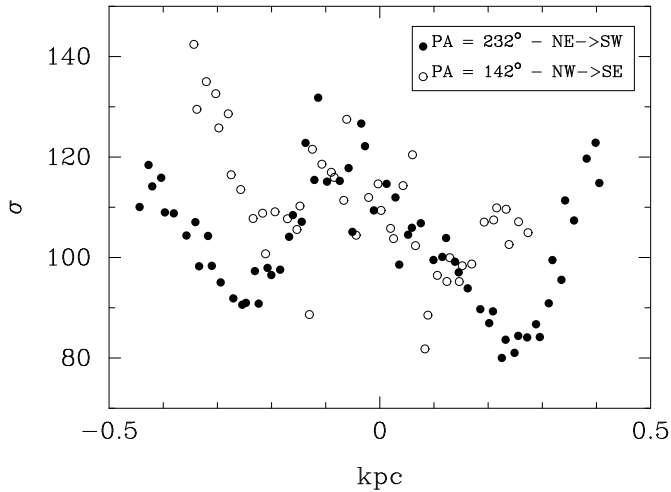}
\caption{One dimensional cuts of the 2D $\sigma$ map of NGC\,2273
within pseudo slits 0\farcs 3 wide crossing the nucleus at PA's
232$^\circ$ and 142$^\circ$.
}
\label{corte-2273}
\end{figure}

The velocity dispersion map (bottom left panel of
Fig. \ref{painel-2273}) presents a ring-like structure with values of
about 60-80\,km\,s$^{-1}$ while in the centre the velocity dispersion
reaches values of up to 100\,km\,s$^{-1}$.
In Fig. \ref{corte-2273} we present one-dimensional cuts of the
velocity dispersion map along
the major and minor axis of the galaxy which helps the visualization
of the ring-like structure.
This structure is co-spatial with the nuclear partial ring
observed in the HST line emission images and color maps of
\citet{FWM00}.
The lower velocity dispersion can be interpreted as a signature of
recently formed stars, which still keep -- at least partially -- the
kinematics of the gas from which they have been formed \citep{woz03}
and thus map the location of a starburst ring with semi-major axis
of $\sim 300$\,pc.

\subsection{NGC\,3227}
\mbox{}

Our radial velocity map (central right panel of Fig.
\ref{painel-3227}) also shows a rotation pattern, where the
redshifted side is to SE of the nucleus, in agreement with the H\,{\sc
i} radial velocity data of \citet{SET00}.
The line of nodes PA obtained from the model ($=
156.7\degr$) agrees within 4$\degr$ with the values found by
\citet{mun95} and \citet{SET00}.

The residuals of the model fitting, shown in the bottom right panel of
Fig. \ref{painel-3227}, are mostly below $\pm\,15$\,km\,s$^{-1}$, 
except for two blueshifted regions approximately along the line of
nodes, which deviate from the model by more than
20\,km\,s$^{-1}$.
The rotation curve (upper right panel) suggests that to NW the 
turnover has been reached already at $\sim 150$\,pc from
the nucleus, while this does not happen to the SE.

\begin{figure}
\centering
\includegraphics[angle=-90,scale=0.45]{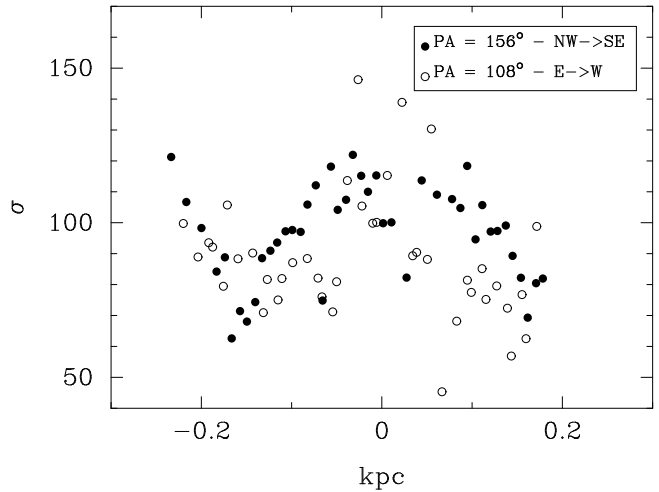}
\caption{One dimensional cuts of the 2D $\sigma$ map of NGC\,3227
within pseudo slits 0\farcs 3 wide crossing the nucleus at PA
$=156^\circ$ and PA $=108^\circ$.
}
\label{corte-3227}
\end{figure}

The velocity dispersion map (bottom left panel of
Fig. \ref{painel-3227}) is characterized by a nuclear region with
$\sigma$ above 100\,km\,s$^{-1}$ which drops in all directions but towards
SE reaching values of $\approx 75$\,km\,s$^{-1}$ or less.
The $\sigma$ values increase again towards E, N and NW as evidenced
in Fig. \ref{corte-3227} by two one-dimensional cuts of the $\sigma$ map.
The loci of low $\sigma$ regions have a good
correspondence to the loci of high CO emission reported by
\citet{SET00} and referred to as a gas ring.
In particular, the loci with the lowest velocity dispersions
correspond to those with highest CO emission.
The loci of low $\sigma$ is also coincident with a ring
of blue $J-K$ colour reported by \citet{cha00}.
Although our FOV seems to be too small to clearly reveal the rise of $\sigma$
beyond the ring in all directions, the coincidence of the low
$\sigma$ loci with both the CO and $J-K$ rings supports also
the existence of a low $\sigma$ ring structure in our data.
As in the case of NGC\,2273 our interpretation is that we are
observing a ring of recent star formation where the low $\sigma$ of
the stars is due to the low velocity dispersion of the cool gas from
which they have formed.

\subsection{NGC\,3516}
\mbox{}

The radial velocity map (central right panel of Fig.
\ref{painel-3516}) shows a rotation pattern which seems to reach the
turnover at the edge of the field, corresponding to a distance of
500\,pc from the nucleus, as can be also seen in the rotation curve
(top right panel of Fig. \ref{painel-3516}).
The velocity field is in good agreement with the 2D
velocity map of \citet{arr97}.
The kinematic
major axis orientation of $\Psi_0 = 48.4\degr$ also
agree with the values found by these authors.
The above value for $\Psi_0$ is also the
orientation of the photometric major axis obtained by fitting
ellipses to the most external isophotes in the i$^{\prime}$-band
acquisition image (corresponding to an ellipse of semi-major axis
$49\arcsec$).

From the rotation curve in the top right panel of
Fig. \ref{painel-3516} we obtain a peak-to-peak amplitude in the
velocity field of $\sim 220$\,km\,s$^{-1}$, significantly larger
than the one measured by \citet{arr97} (160\,km\,s$^{-1}$).
This discrepancy can be understood if we recall that those authors
applied a 1\farcs5 smoothing to the data, which would flatten the
rotation curve.
We note that the peak-to-peak amplitude of the stellar
rotation is well below the one obtained from the gas kinematics
(592\,km\,s$^{-1}$), by \citet{mul92}.

Our velocity dispersion map (bottom left panel of
Fig. \ref{painel-3516}) shows typical values in the central region
larger than 150\,km\,s$^{-1}$
surrounded by a region with lower $\sigma$ (as low as $\sim
90$\,km\,s$^{-1}$).
From measurements of the Mg\,{\sc i}\,b absorption band,
\citet{arr97} reported a central $\sigma$ of $164 \pm
35$\,km\,s$^{-1}$ (within an aperture of 3\arcsec), which is
consistent with our data, but a 2D distribution of line
widths does not show any clear structure.
In our data we can see lower $\sigma$ values at locations
surrounding the kinematic major axis direction at an average
distance from the nucleus of $\sim 400$\,pc.

The residuals from the radial velocity fit (bottom right panel of
Fig. \ref{painel-3516}) are mostly within the adopted error of
$\pm\,15$\,km\,s$^{-1}$.

\subsection{NGC\,4051}
\mbox{}

In this galaxy the nuclear continuum and broad Ca-T emission lines
preclude reliable measurements of the stellar kinematics using the
Ca-T absorption lines within 0\farcs 5 from the nucleus.
This region has thus been masked for the model fitting.

The radial velocity map (central right panel of Fig.
\ref{painel-4051}) is dominated by the typical rotation pattern
observed in all galaxies so far. 
The rotation curve (upper right panel of Fig. \ref{painel-4051}) has
a peak-to-peak amplitude of only 130\,km\,s$^{-1}$ for an
inclination of the galaxy $i \sim 41\degr$ (assuming
$i = \cos ^{-1} b/a$ where $a$ and $b$ are the major and minor axis
photometric lengths obtained from NED) and the turnover seems to
occur very close to the nucleus, at $R\sim$50\,pc.

The residuals from the model fit are within the
$\pm\,10$\,km\,s$^{-1}$ error limits (bottom right panel of Fig.
\ref{painel-4051}) and do not show any well defined structure.

The velocity dispersion map (bottom left panel of Fig.
\ref{painel-4051}), is also  irregular with $\sigma$ values ranging
from a minimum of $\sim 60$\,km\,s$^{-1}$ to a maximum of $\sim
100$\,km\,s$^{-1}$.
Nevertheless it was not possible to extend the measurements closer
to the nucleus than $\sim 0\farcs5$, making it difficult to draw any
conclusion on the $\sigma$ distribution.

\subsection{NGC\,4593}
\mbox{}

For this galaxy, as for NGC\,4051, the continuum from the Seyfert
nucleus precludes reliable measurements of the stellar kinematics in
the central region.

The radial velocity map (central right panel of Fig. \ref{painel-4593})
shows a clear rotation pattern in which the kinematical major axis
runs approximately along E-W with the redshifted side to the E.

The peak-to-peak amplitude of the velocity field is
255\,km\,s$^{-1}$, but the rotation curve (top right panel of 
Fig. \ref{painel-4593}) suggests that the maximum velocity may occur
beyond the edges of the observed FOV, which corresponds to $\sim
700$\,pc from the nucleus along the kinematic major axis.

\begin{figure}
\centering
\includegraphics[angle=-90,scale=0.45]{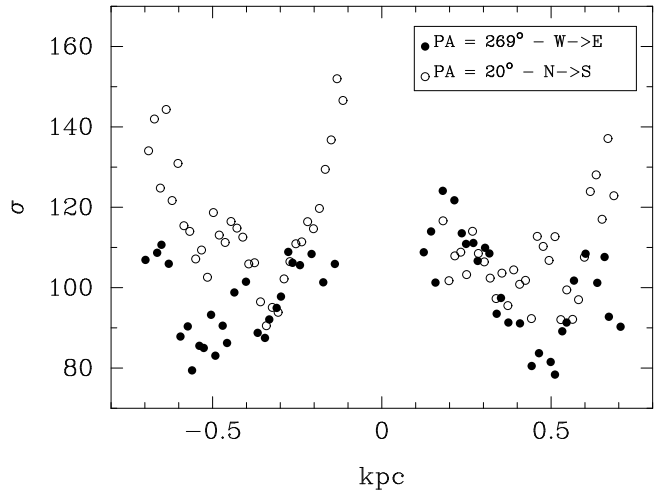}
\caption{One dimensional cuts of the 2D $\sigma$ map of NGC\,4593
within pseudo slits 0\farcs 3 wide crossing the nucleus at PA
$=269^\circ$ and PA $=20^\circ$.
}
\label{corte-4593}
\end{figure}

The velocity dispersion map (bottom left panel of Fig.
\ref{painel-4593}) hints on a rise of the observed values towards
the nucleus.
At the most central but usable region we find values as high as
120\,km\,s$^{-1}$.
At $\sim 2\farcs 5$ from the nucleus ($\sim 400$\,pc), the measured
values drop to 80-90\,km\,s$^{-1}$, most notably along the
kinematic major axis as in the case of NGC\,3516.
In the directions approximately perpendicular to the above the
velocity dispersion is found to decrease slower, then rise again,
reaching up to 130\,km\,s$^{-1}$ at the edges of the field.
The map thus hints at the presence of a partial
ring structure also for this galaxy.
This is supported by the one-dimensional cuts along the major and
minor axis of the velocity dispersion map shown in Fig.
\ref{corte-4593}.
Alternatively this region of low $\sigma$ values can be associated
with a tightly wound spiral arm which can be observed in an HST
F547M image from \citet{MGT98}.

\citet{gon97} have reported the presence of a broken starburst ring
in an H$\alpha$ narrow band imaging.
But a comparison of our data with theirs shows that our low $\sigma$
partial ring is nevertheless internal to their H$\alpha$ ring.

The radial velocity residuals (bottom right panel of Fig.
\ref{painel-4593}) are within the $\pm\,15$\,km\,s$^{-1}$ error
limits, but show some resemblance to the $\sigma$ map, suggesting
systematic departures from circular rotation in association with the
apparent partial ring of lower velocity dispersion.

\subsection{NGC\,4941}
\mbox{}

A rotation pattern is again observed in the stellar kinematics
(central right panel of Fig. \ref{painel-4941}).
The rotation curve (top right panel of Fig. \ref{painel-4941})
indicates that the maximum amplitude is located beyond the observed
region, which corresponds to a radial distance from the nucleus of
only 200\,pc for this galaxy.

The velocity dispersion map (bottom left panel of Fig.
\ref{painel-4941}) is almost flat within $\sim$ 100 pc from the
nucleus, with central values $\sim$ 135 km\,s$^{-1}$, decreasing to $\sim$
100 km\,s$^{-1}$ at the edges of the field ($\sim$ 200 pc).

The radial velocity residuals (bottom right panel of Fig.
\ref{painel-4941}) are within the $\pm\,15$ km\,s$^{-1}$ error limit
and show no obvious pattern.

\section{Summary and Conclusions}
\mbox{}

In this work we have obtained 2D maps of the 
stellar kinematics of the inner few hundred parsecs of 6 nearby
Seyfert galaxies at sub-arcsecond angular resolution, corresponding
to spatial resolutions ranging from 30 to 180 parsecs at the
galaxies. 

The stellar velocity field is dominated by rotation, well
represented by a simple model where the stars follow plane
circular orbits under a Plummer potential.
The residuals between measured and modeled radial velocities are
always smaller than $\pm\,25$\,km\,s$^{-1}$.

In most cases, the turnover of the rotation curve seems to occur
within or close to the edges of the observed field, at radial
distances ranging from $\sim 50$\,pc for NGC\,4051 to $\sim
500-700$\,pc for NGC\,3516 and NGC\,4593.
Only for NGC\,4941, for which our observations reach only as far as
$\approx 200$\,pc from the nucleus, the turnover was not observed within the
IFU FOV.

The case of NGC\,4051 is particularly interesting
because the turnover is at only $\sim 50$\,pc 
from the nucleus,
suggesting that the stellar motions are dominated by a highly
concentrated gravitational potential.
Indeed, the scale length of the Plummer potential obtained for this
galaxy is the smallest of the the sample, $A = 31$ pc.
Adopting $\sigma = 85$\,km\,s$^{-1}$ for the
nuclear velocity dispersion (the mean value of innermost usable pixels),
and the \citet{tre02} relation, we infer a black hole mass of
$4.5 \times 10^6$ M$_\odot$, which implies a sphere of influence of
only $2.5$\,pc radius.
These values are consistent with those by \citet{pet04}, who
obtained a black hole mass of $1.9\times 10^6$ M$_\odot$
corresponding to a sphere of influence of $1.2$\,pc radius.
Such concentrated potential is unlikely to be responsible for the
small potential-scale size we have found.
A more likely interpretation is that the bulge is itself compact and
concentrated.
Indeed, unlike most Seyferts, this galaxy is late type (Scd),
with a small bulge effective radius.

The velocity dispersion maps are featureless for NGC\,4051 and
NGC\,4941.
For the other galaxies the velocity dispersion maps show higher values
at the centre and smaller values at $200-400$\,pc from the nucleus.
In the case of NGC\,3516 the lowest $\sigma$ values are observed
towards the border of the IFU field, at location close to the
kinematic major axis.
For NGC\,2273, the spatial distribution of lower
$\sigma$ values defines a ring-like morphology which is co-spatial
with partial nuclear ring structures seen in line emission images
and color maps of previous works.
For NGC\,3227, the velocity dispersion map shows a very good
correspondence between the region of low $\sigma$ values -- which
seems to delineates a partial ring structure -- and regions
of high $^{12}$CO (2-1) emission and blue $J-K$ colours observed in
previous works.
A partial ring of low $\sigma$ values is also hinted in NGC\,4593.
As for the nuclear $\sigma$-drops found by previous authors (e.g.
\citealt{ems01,gar99,mar03}) the circumnuclear $\sigma$-drops
we found in the present study can be interpreted as regions of
higher (past or present) gas concentration, harboring younger stars
which still preserve the lower velocity dispersion of the original
gas from which they have formed.

\section*{Acknowledgments}
\mbox{}

We thank the referee, Eric Emsellem, for useful suggestions which
helped to improve the paper.
We acknowledge support from the Brazilian funding agencies CNPq and
CAPES, the US Naval Research Laboratory, where basic research is
supported by the Office of Naval Research.

The PI is very grateful to Dr. Roberto Maiolino, Dr. Alessandro
Marconi and the Arcetri group for the hospitality during his
permanence at the Observatory of Arcetri where part of the work has
been developed.

We thank also Natalia Vale Asari for helpful discussions.

Based on observations obtained at the Gemini Observatory,
which is operated by the 
Association of Universities for Research in Astronomy, Inc., on 
behalf of the international Gemini partnership of Argentina, 
Australia, Brazil, Canada, Chile, the United Kingdom, and the 
United States of America (observing programmes GN-2002B-Q-15,
GN-2003A-Q-20, and GN-2004A-Q-1).

We have used the Levenberg-Marquardt non-linear least squares
algorithms from M.I.A. Lourakis available from
http://www.ics.forth.gr/$\sim$lourakis/levmar/.

\bsp

\label{lastpage}

\end{document}